# Enhancing Patent Retrieval using Text and Knowledge Graph Embeddings: A Technical Note

L. Siddharth[1], Guangtong Li[*,1], Jianxi Luo[1]

[1]*Data-Driven Innovation Lab, Engineering Product Development Pillar,*

*Singapore University of Technology and Design (SUTD), 8 Somapah Road, Singapore 487372*

**Abstract**

Patent retrieval influences several applications within engineering design research, education, and practice as well as applications that concern innovation, intellectual property, and knowledge management etc. In this article, we propose a method to retrieve patents relevant to an initial set of patents, by synthesizing state-of-the-art techniques among natural language processing and knowledge graph embedding. Our method involves a patent embedding that captures text, citation, and inventor information, which individually represent different facets of knowledge communicated through a patent document. We obtain text embeddings using Sentence-BERT applied to titles and abstracts. We obtain citation and inventor embeddings through TransE that is trained using the corresponding knowledge graphs. We identify using a classification task that the concatenation of text, citation, and inventor embeddings offers a plausible representation of a patent. While the proposed patent embedding could be used to associate a pair of patents, we observe using a recall task that multiple initial patents could be associated with a target patent using mean cosine similarity, which could then be utilized to rank all target patents and retrieve the most relevant ones. We apply the proposed patent retrieval method to a set of patents corresponding to a product family and an inventor's portfolio.

---

[*] Corresponding author.
Email: guangtong_li@mymail.sutd.edu.sg

# 1. Motivation

A patent is intellectual property granted for an artefact (product or process)[1]. A patent document is identified using fields such as patent number, inventor, assignee, classifications, citations, etc., as well as organised according to sections such as title, abstract, background, summary, technical field, description of embodiments/drawings, examples, claims, etc. A set of patent documents, for example, corresponding to an inventor, having a common assignee (e.g., Valspar Sourcing), or adhering to a common theme (e.g., curable coating compositions), etc., could partially be considered a representative of a domain of knowledge.

Since the development TRIZ [1], patent documents have enabled various applications in engineering design research [2], education [3], and practice [4], while also aiding business applications concerning innovation [5], intellectual property and knowledge management [6]. For instance, domain concepts extracted from patent documents, such as a range of terms [7], topics [8], facts [9], are useful in design-cum-learning environments, form building blocks for domain ontologies [10]–[12], and also indicate TRIZ trends for technology forecasting [13]. In view of business applications, studying the emergent properties of citation and inventor networks offers insights that are useful for bibliometric [14] and management studies [15]. In addition to these applications, Jiang et al. [16] provide a comprehensive review of scholarly contributions in engineering design that utilise patent documents.

The above-stated applications primarily require a relevant set of patents, retrieving which is a challenge despite the existence of various patent search tools, e.g., PatentsView[2] and InnoGPS[3] [17]. While such resources are suitable for retrieving patents w.r.t., fields (e.g., inventor id), the patents retrieved using search keywords largely include false positives. Human intervention is often necessary to identify the relevant patents and reformulate search keywords. According to design literature, keyword reformulation is supported by various means that involve identifying troponyms [18], antonyms [19], relevant concepts [20], domain vocabulary [21] as well as channelling keywords into categories like Function, Behaviour, and Structure [22].

---

[1] https://www.wipo.int/about-ip/en/
[2] https://patentsview.org/apis/api-query-language
[3] https://www.innogps.com

Although the above-stated supports exist, keyword-based retrieval often results in several false positives, mainly due to the usage of a keyword (e.g., 'storage') in various contexts that are not relevant to the application. To mitigate this issue, it is preferable to retrieve patents using a sample of patents as input. Let us consider the patent[4] titled "Method of coating a packaging container using cross-linkable polyester-polyurethane" as an example. Upon searching for the concepts mentioned in the title, we identify that various patents use polyester-polyurethane for coating, while only one other patent[5] – "Polymer having polycyclic groups and coating compositions thereof" uses the material in packaging containers.

Retrieving patents from an initial set of patents (e.g., inventor's portfolio) could therefore result in a lesser number of false positives compared to keyword-based retrieval. Although design literature includes various supports for keyword formulation and expansion, contributions that focus on patent-to-patent retrieval seem less apparent. The objective of our research is therefore to devise a method for retrieving patents related to an initial set of patents. The method is expected to contribute to various design applications and some business applications in both academia and industry.

---

[4] https://patents.google.com/patent/US6730361B2/en?oq=US6730361B2
[5] https://patents.google.com/patent/US8946316B2/en?oq=US8946316B2

## 2. Background

Patent retrieval, as indicated in Figure 1, often requires a vector representation or embedding method for all patents so that cosine similarity could be used to measure the association strength between an initial patent and a target patent. Such a method also must inform how multiple patents in the initial set could be associated with each target patent so that all target patents could be ranked based on the overall strength of association. Subsequently, depending on the number of desired patents, the target patents ranked above shall be retrieved.

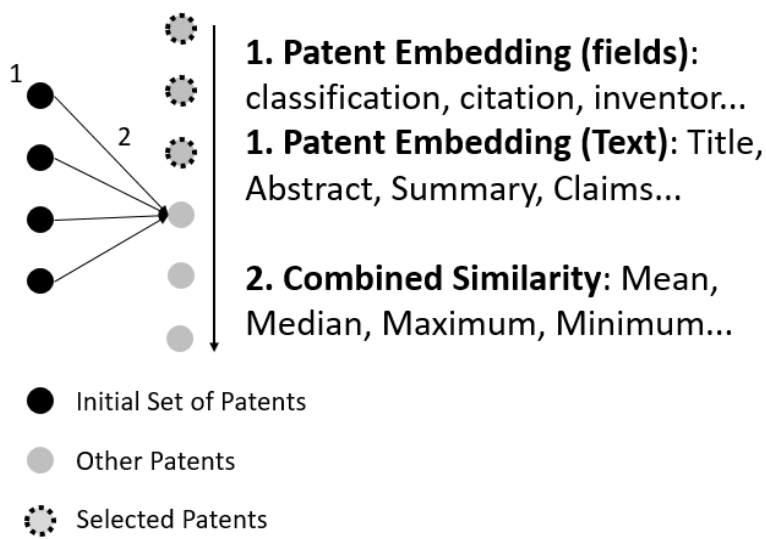

Figure 1: Patent-to-Patent retrieval.

A basic approach to obtaining vector representation of a patent is through the indexing given by the classification codes, e.g., 'F02' [23]. For instance, Luo et al. [24] locate patents or a patent portfolio in the total technology space according to their position vectors whose dimensions correspond to international patent classes (IPC). Since several patents could share similar classifications, as an alternative, Murphy et al. [25] index 2,75,000 patents using 1700 function terms that were obtained by analysing 65,000 patents. Song and Fu [26] index 1,060 patents using heuristically developed terms that correspond to component, behaviour, or material. Wodehouse et al. [27] extract adjectives and keyphrases from patents using NLTK[6] and RAKE[7] respectively. Upon computing semantic relatedness between adjectives and keyphrases, they index patents using adjective-keyphrase pairs.

---

[6] https://www.nltk.org/howto/sentiment.html
[7] https://pypi.org/project/rake-nltk/

As indexing terms could carry multiple senses across domains, it is necessary that a vector representation captures the context that is communicated within the patent text as well as the patent fields. Scholars have therefore adopted approaches like bag-of-words [28] and Word2Vec [29] for embedding patent text onto a vector representation in a high-dimensional vector space. Upon extracting key phrases from patent titles using KeyBERT[8], Zhu and Luo [30], [31] fine-tune GPT-2[9] (a large pre-trained transformer from OpenAI) using keyword-title pairs, learning embeddings for patent titles in the process. They subsequently propose using keywords as prompts for the fine-tuned GPT-2 to generate design idea descriptions mincing the syntax of patent titles.

Jiang et al. [32] obtain embeddings for 0.8 million patents that incorporate engineering drawings, text, and citations. Upon disintegrating a drawing into simpler images using YOLOv3[10], they obtain feature vectors for these using VGG-19[11]. They train Dual-VGG (based on Convolutional Neural Networks - CNN) to map patent drawings to patent classification codes such as A to H [33]. From titles and abstracts, they extract technical terms and their embeddings through TechNet[12] [34]. They apply an attention mechanism (based on Recurrent Neural Networks - RNN) to integrate these terms embeddings onto sentence-level and subsequently to patent-level text embeddings.

Once drawing-based and text-based embeddings are learnt, they concatenate these to train another RNN for mapping patent (text and drawing) to the classification codes. They then represent patents in the citation network using text-cum-drawing embeddings to train GraphSAGE (based on Graph Neural Networks - GNN) for classifying patents in the network according to one or more classification codes. The resultant patent embeddings thus integrate drawings, text, and citations.

Thus far, the embeddings obtained by Jiang et al. [32] offer the most comprehensive representation of a patent. However, engineering drawings captured in their embeddings, are merely coarse depictions of certain embodiments in which, the components may not carry sufficient features for recognizing these as distinct objects[13]. Such components carry specific labels that are numbered in the drawings and referred to in the main body of text. In view of a patent document, a drawing depicts a configuration that is better understood as a set of specific components and is less significant when captured without the description offered by the text.

---

[8] https://github.com/MaartenGr/KeyBERT
[9] https://openai.com/blog/tags/gpt-2/
[10] https://viso.ai/deep-learning/yolov3-overview/
[11] https://www.mathworks.com/help/deeplearning/ref/vgg19.html;jsessionid=a31a03d426b5e174845c9332c7de
[12] http://www.tech-net.org/

The text and citations captured by Jiang et al. [32], however, constitute significant facets of knowledge. Using a case study on self-propelled spherical rolling robots, Song and Luo [35] demonstrate that text (title and abstract) along with citations and inventors guided the retrieval of 150 patents from an initial set of 23 patents retrieved solely by keyword-based heuristic searches. Our method, as proposed in the following section, captures text, citations, and inventors using preferred embedding techniques from natural language processing and knowledge engineering literature.

---

[13] An engineering drawing shall not be treated equivalent to images in which objects could be detected using common-sense entities (e.g., aircraft, elephant). The components depicted in an engineering drawing carry specific labels that are not part of common-sense ontologies utilized by the object-detection techniques.

# 3. Method

## 3.1. Source Data Processing

Using an open-source data platform called PatentsView, we gather title-abstract, citations, and inventors for 7,988,799 utility patents granted by USPTO from 1976 to 2021. Once gathered, as illustrated in Figure 2, we concatenate the title and abstract into a single sequence. We encode the citation and inventor information from adjacency tables into triplets or facts such as <patent, *cite*, patent> and <inventor, *write*, patent> and store these as citation and inventor knowledge graphs, which are visualized in Figure 2. We apply Sentence-BERT and TransE methods to obtain embeddings of text and knowledge graphs respectively. We build a classifier using these embeddings and identify the suitable approach to integrate these embeddings. In the following subsections, we elaborate on these steps.

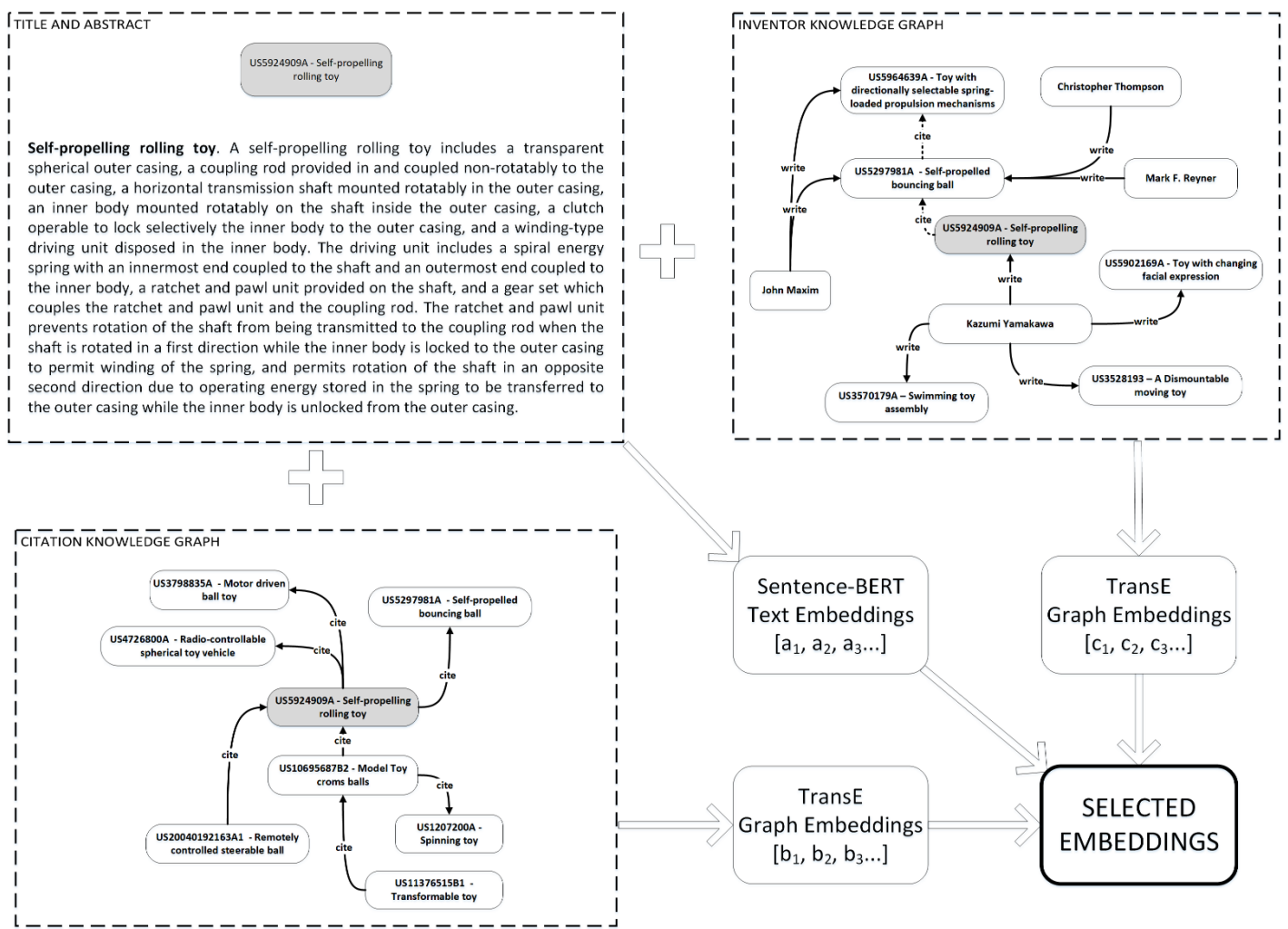

Figure 2: Outline of the method using an illustrative example.

## 3.2. Text Embeddings

Our method requires an embedding technique for title and abstract concatenated into a piece of text as indicated in Figure 2. BERT[14] is among state-of-the-art language representation models that are trained on the sentences found in Wikipedia (2.5 billion words) and BookCorpus (800 million words) for masked-language modelling and next-sentence prediction. BERT has been fine-tuned and adopted to support various engineering design applications, e.g., to summarize engineering documents [36]. Upon providing a sentence as input, the pre-trained BERT model primarily converts the input into a sequence of tokens as shown in the example below[15].

*"A sliding switch 106 is positioned between rims 104 of spool 96"*

*['[CLS]', 'a', 'sliding', 'switch', '106', 'is', 'positioned', 'between', 'rim', '##s', '104', 'of', 'sp', '##ool', '96', '[SEP]']*

In the above example, the sentence is first separated using white space and the resulting tokens are dissected (e.g., 'rims' → 'rim' + '##s') such that all tokens in the generated sequence are recognized as part of BERT vocabulary. BERT includes [CLS] and [SEP] tokens to indicate either end of the sequence. Upon taking the sequence as input, the pre-trained BERT model provides contextualized embeddings for each token, i.e., the embedding for 'switch' in the above sequence would be different from that of the same token in a different sentence. Contextualized embeddings take into account word-sense disambiguation, which is absent in word embedding models that are utilized by Jiang et al. [32] in their approach.

To obtain the embedding of the whole sequence, it is a common approach to use the pooled output (mean or max of token embeddings) given by BERT or the embedding of the [CLS] token. The sequence embeddings could be used in tasks like classification, sentiment analysis, clustering etc. We require an embedding technique for a patent such that cosine similarity between a pair of patents represents a meaningful strength of association between these. Despite offering adequate performance in other tasks (e.g., classification), the sequence embeddings given by BERT were not found to return meaningful similarity between a pair of sequences [37].

---

[14] Bidirectional Encoder Representations from Transformers - https://github.com/google-research/bert
[15] https://patents.google.com/patent/US5297981A/en?oq=US5297981A

We therefore use a variant of BERT, namely Sentence-BERT, which was trained using a modified BERT network, namely Siamese network structure [37], for predicting similarity scores between a pair of sentences. In our work, we directly utilise the pre-trained Sentence-BERT model[16] for obtaining embeddings of sequences (of length < 256) containing titles and abstracts. The output of Sentence-BERT, an embedding of size 384 (denoted as $[a_1, a_2, a_3 ... a_{384}]$ in Figure 2), represents the primary facet of knowledge communicated in patent document.

### 3.3. Knowledge Graph Embeddings

Our method (as outlined in Figure 2) also requires an embedding technique for capturing citation and inventor information onto patent representation. Jiang et al. [32] capture citation information using GNNs, wherein, the embedding of a patent, which is a node in the citation network, is learnt only based on its neighbourhood. Such an approach does not capture the structure and semantics of the entire citation or inventor network, which are necessary for associating patents that indirectly connected. We therefore adopt a knowledge graph embedding technique that learns patent embedding based on the graph as a whole.

A knowledge graph is a scheme for storing and visualizing a set of facts, each represented as $\langle h, r, t \rangle$ where h – head entity, r – relation, and t – tail entity. As illustrated in Figure 2, various facts corresponding to the types <patent, *cite*, patent> and <inventor, *write*, patent> could be integrated and visualized in a graph format. Such a graph could be embedded into vector spaces using embedding algorithms in the knowledge engineering literature. These algorithms could broadly be divided into tensor decomposition (e.g., DistMult, ComplEx), geometric (e.g., TransE, RotatE), and deep learning (e.g., ConvE, CapsE) based algorithms.

A general procedure to train the above-stated models is the following: 1) random initialization of embeddings for all entities and relations, 2) corrupt or negative facts ($\{\langle h', r, t \rangle\}$ or $\{\langle h, r, t' \rangle\}$) for each correct or positive fact - $\langle h, r, t \rangle$, 3), and 3) a triplet loss function $\emptyset(h, r, t)$ that is specific to an embedding algorithm. While training, the embeddings are learnt such that the loss function is minimized for positive facts and maximized for negative facts [38]. The embeddings thus learnt, could be applied to link prediction or domain-specific tasks for comparing the performances of different embedding algorithms.

---

[16] https://github.com/UKPLab/sentence-transformers

In our earlier work [39], we trained seven embedding algorithms – TransE (L1 and L2), TransR, RESCAL, DistMult, ComplEx, and RotatE on a knowledge graph built using patent metadata (e.g., <patent, *cite*, patent>). Among these, TransE – L2 exhibited superior performance in link prediction and domain-specific tasks. In this work, therefore, we use TransE for embedding purposes. TransE is a geometric-based algorithm, wherein, the triplet loss function $\emptyset(h, r, t)$ is a distance $d$ calculated between the transformation $\tau(h, r) = h + r$ of $h$ - head embedding through $r$ - relation embedding and the $t$ – tail embedding [40]. The objective of training TransE is to minimize and maximize the distance $d(h + r, t)$ for positive and negative facts respectively.

We apply the TransE model to individually embed the citation and inventor graphs (see Figure 2) using the graph learning library DGL-KE[17] and the required hardware[18]. To be consistent with the text embeddings given by Sentence-BERT, we set the embedding size as 384 for training TransE as well. The resultant embeddings of patents (denoted as $[b_1, b_2, b_3 ... b_{384}]$ and $[c_1, c_2, c_3 ... c_{384}]$ in Figure 2) from citation and inventor knowledge graphs, individually embody other facets of knowledge in a patent document, alongside the text embeddings (denoted as $[a_1, a_2, a_3 ... a_{384}]$ in Figure 2).

**3.4. Embedding Selection**

Upon application of Sentence-BERT and TransE models, we have obtained text, citation, and inventor embeddings. In this section, using the performance in a classification task, we identify which among these or which combinations of these are best suited for patent representation. Each patent is assigned one or more classification codes (e.g., F02) that are given by the Cooperative Patent Classification (CPC) system[19]. Using this dataset, we train a classifier that inputs a patent embedding of a given size, includes a hidden layer of size 384, includes a dropout layer (dropout rate = 0.1), and outputs a vector of size 130, which corresponds to the number of classification codes in the CPC scheme.

We train the classifier with different types of inputs, i.e., patent embeddings of types: A, B, C, A + B, B + C, A + C, A + B + C, [A, B], [B, C], [A, C], [A, B, C]; wherein A, B, C denote text, citation, and inventor embeddings respectively and [*, *] and * + * denote concatenation and addition operations respectively. For training, we use the following arrangement: *Dataset Distribution*: 8:1:1, *Optimizer*: Adam (learning rate: 5e-5), *Loss Function*: Binary Cross Entropy, *Batch Size*: 16, *Training loops*: 10.

---

[17] DGL-KE – Deep Graph Library for Knowledge Graph Embedding: https://aws-dglke.readthedocs.io/en/latest/commands.html
[18] We conduct training using a server with the following configuration: 8 x NVIDIA Tesla P100-SXM2-16G GPUs and 512 GB of Memory.
[19] https://www.uspto.gov/web/patents/classification/cpc/html/cpc.html

We report the performance of the classifier in Table 1 using the following metrics: TOP-n accuracy, Precision, Recall, and F-Score. TOP-n accuracy considers a prediction as correct if one of the classification codes is correctly identified among the n top-ranked codes. Since patents are usually assigned more than one classification code, TOP-n accuracy is suitable for this classification task. Table 1 shows that [A, B, C], which is the concatenation of A, B, and C, offers the best classification performance. The observation is consistent with Song and Luo [35] that a combination of text, citation, and inventor information aids patent retrieval.

Table 1: Performance of different embeddings on a classification task, where A – Text Embedding, B – Citation Embedding, and C – Inventor Embedding

| Patent Embedding Type | Patent Embedding Size | Top 1-Accuracy ↓ | Top 5-Accuracy | Top 10-Accuracy | Precision | Recall | F1 |
|---|---|---|---|---|---|---|---|
| [A, B, C] | 1152 | **0.702** | **0.947** | **0.979** | **0.659** | 0.812 | **0.726** |
| [A, B] | 768 | 0.7 | 0.947 | 0.978 | 0.654 | **0.813** | 0.723 |
| A + B | 384 | 0.684 | 0.936 | 0.972 | 0.634 | 0.791 | 0.702 |
| A + B + C | 384 | 0.662 | 0.923 | 0.966 | 0.611 | 0.767 | 0.678 |
| [B, C] | 768 | 0.656 | 0.914 | 0.958 | 0.617 | 0.757 | 0.678 |
| B | 384 | 0.654 | 0.912 | 0.956 | 0.614 | 0.754 | 0.675 |
| [A, C] | 768 | 0.638 | 0.907 | 0.956 | 0.593 | 0.748 | 0.66 |
| A | 384 | 0.637 | 0.905 | 0.955 | 0.591 | 0.742 | 0.656 |
| B + C | 384 | 0.617 | 0.885 | 0.941 | 0.584 | 0.708 | 0.638 |
| A + C | 384 | 0.601 | 0.882 | 0.942 | 0.562 | 0.704 | 0.624 |
| C | 384 | 0.175 | 0.434 | 0.581 | 0.443 | 0.145 | 0.218 |

Such a combination could be achieved through concatenating ([A, B, C]), adding (A + B + C), stacking (A → B → C), and others (A + B → C). Among these, we only considered concatenation and addition because we obtained individual embeddings through exclusive training approaches. In a different approach, Jiang et al. [32] learn the text and drawing embeddings individually before passing these as inputs to GNN where citation information is harnessed. While such approaches might accumulate errors[20] through stacking, only the concatenation approach distinctly captures all facets of a patent representation.

The embeddings that exhibit slightly lower performance than [A, B, C], involve at least a combination of text and citation, preferably through concatenation. As we have illustrated in the inventor knowledge graph of Figure 2, the patents "self-propelled rolling toy" and "self-propelled bouncing ball" are not connected (indicated by a dashed arrow) through an inventor but through citation. It is therefore necessary to capture both citation and inventor, preferably through concatenation than through addition as indicated by the higher performance of [B, C] compared to B + C.

### 3.5. Combined Similarity

As the concatenation of text, citation, and inventor embeddings exhibit the best performance in the classification task, we use the same for patent representation henceforth. While the chosen embedding type could be used to associate a pair of patents, a patent retrieval (as outlined in Figure 1) requires associating multiple initial patents and a target patent so that all target patents could be ranked for similarity or relevance. Such an association could be performed in many ways including, as considered in our work: maximum, minimum, mean, and median.

To identify the best-preferred way to perform many-to-one association of patents, we examine the coverage within a set of patents identified by Song and Luo [35]. To retrieve patents related to self-propelled spherical rolling robots, they found 10 patents through keyword search (e.g., through "rolling toy" and "rolling robot") and expanded the set to 109 patents[21] upon examining citations and inventor links. Beginning with 10 patents given by Song and Luo [35], we rank the remaining 99 patents among other patents in the database using combined (through mean, median, min, or max) cosine similarity.

We sort the ranks corresponding to 99 patents in ascending order for each approach and plot the logarithm of ranks in Figure 3. In an ideal scenario (indicated as baseline in Figure 3), among all remaining patents, the 99 patents identified by Song and Luo [35] should be ranked as top 99. Since such a scenario is highly unlikely, the preferred approach (mean, median, min, and max) should rank 99 patents as above as possible. In Figure 3, such an approach should hence exhibit the lowest Area Under the Curve (AUC). We identify that the rank plot for 'mean' exhibits the lowest AUC. It indicates that the usage of mean cosine similarity to associate the initial set of patents and a target patent enables better retrieval of patents.

---

[20] We could not verify the performance of Jiang et al. [32] through comparison in Table 1 because their dataset is restricted to 0.8 million patents alone.
[21] The actual number of patents retrieved via keyword search is 23 and then expanded to 150. Since these sets of patents include those that were granted before 1976, we consider 10 patents (keyword search) and 109 patents (upon expansion) in our evaluation.

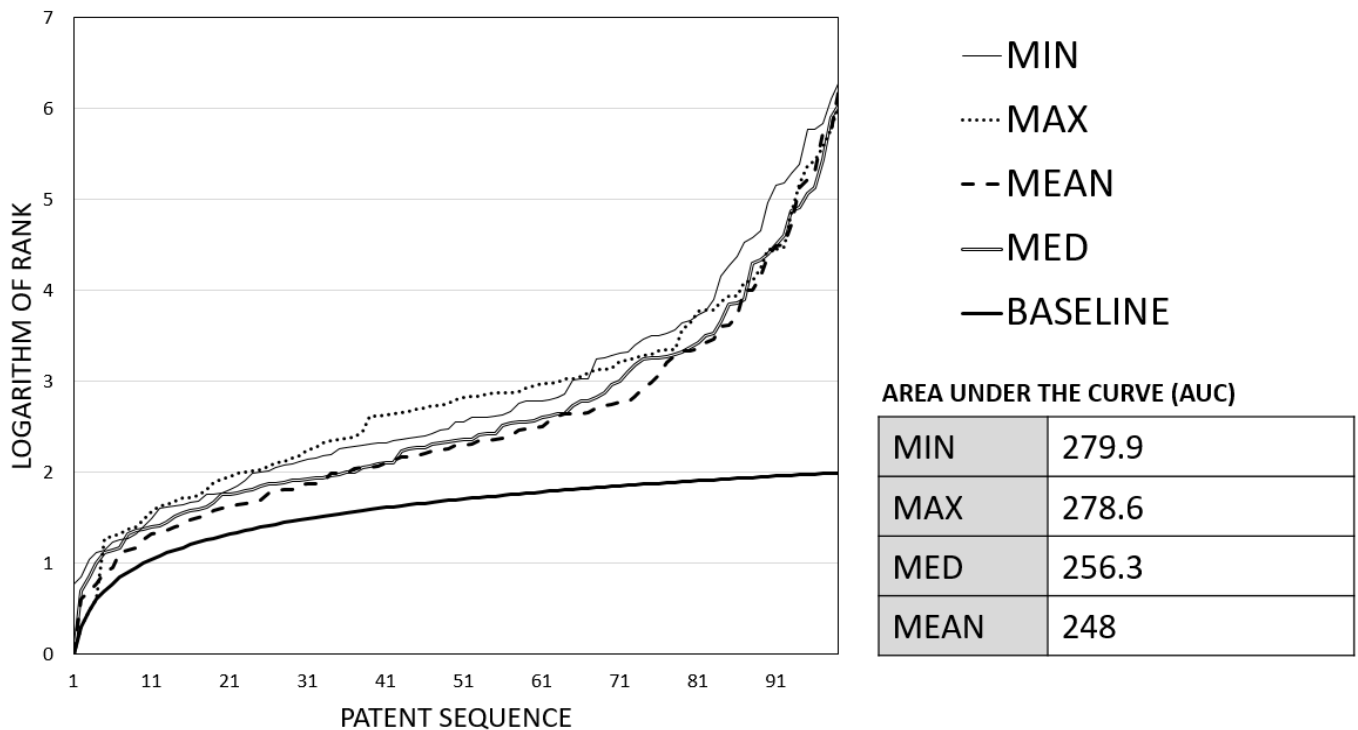

Figure 3: The ranks assigned to 99 patents of Song and Luo [35] using min, max, median and mean.

# 4. Application

In this section, we apply the proposed method to a couple of sets of patents to retrieve related patents. In the first application, we use the 10 patents identified by Song and Luo [35] through keyword search (using the terms "rolling toy" and "rolling robot") for populating patents related to self-propelled spherical rolling robots. In the second application, we begin with an inventor's portfolio of patents and retrieve patents that are similar to the portfolio. We have uploaded the patent numbers, titles, abstracts, dates etc., as a CSV[22] for both sets of patents. For both applications, we compare the initial set and retrieved set of patents using the concepts extracted from these.

## 4.1. Product Family (Design)

We use 10 patents identified by Song and Luo [35] to retrieve 36 related patents. The purpose of this application is not to compare the retrieved set of patents with the 99 patents retrieved by Song and Luo [35], but to understand the advantage of the retrieved set over the initial set in terms of offering useful concepts of inspiration for utility in the design process.

Among the retrieved 36 patents, a few are ranked above (e.g., 4363187[23]) despite the absence of keywords "rolling robot" or "rolling toy" and not sharing any citation or inventor with the initial set of patents. While such observation is unlikely through query-based retrieval methods (e.g., PatentsView), it could be attributed to a couple of aspects. First, TransE captures the structure and semantics of the entire citation/inventor graph so that a pair of patents sharing similar neighbourhood structure could return higher cosine similarity. Second, Sentence-BERT, which was trained upon BERT, additionally has the ability to compute a meaningful similarity between sentences despite having no common terms.

To understand the significance of retrieved patents, we extract the concepts (as noun phrases) from retrieved patents and compare these to that of the initial set of patents. We use spaCy[24] to extract the noun phrases from the titles and abstracts of both sets of patents and select the most frequent phrases. To avoid repetition of concepts, 1) we convert the phrases to lower case; 2) remove the determinants (e.g., the, a, an); 2) remove the count terms (e.g., first, second); 3) select phrases that include more than one word. Upon filtering, in Figure 4a, we present the concepts (including their frequencies) for both sets of patents.

---

[22] https://github.com/siddharthl93/patent-retrieval/blob/12097718f879d639a5124cc3b96014185a51bbc7/patent-list-public.csv
[23] Toy capable of repeatedly upsetting and then righting itself - https://patents.google.com/patent/US4363187A/
[24] https://spacy.io/usage/linguistic-features

| KEYWORD SEARCH PATENTS (N = 10) | RETRIEVED PATENTS (N = 36) |
|---|---|
| outer casing (19), inner body (12), coupling rod (9), ratchet and pawl unit (6), magnetic element (5), coupling ring (5), **rolling toy (4), self-propelling rolling toy (4)**, sphere body (3), gear assembly (3), support member (3), inner toy body (3), pivot axis (3), system controller (3), magnetic switch (2), transparent spherical outer casing (2), horizontal transmission shaft (2), clutch operable (2), winding-type driving unit (2), driving unit (2), spiral energy spring (2), innermost end (2), outermost end (2), opposite second direction (2), operating energy (2), rolling movement (2), plate portion (2), horizontal surface (2), inner surface (2), **autonomous rolling robot (2), hollow sphere (2)**, geometric center (2), remote signal receiver (2), electric motor (2), power source (2), more sensors (2) | toy vehicle (14), system controller (9), driving mechanism (6), more sensors (6), **spherical housing (6), rolling member (6), spheroid shell (5)**, stator portion (5), drive assembly (4), pivoting arm (4), spherical cap (4), more wheels (4), central housing (4), reversible motor (4), **shell parts (4), robot ball (4)**, motorized mechanism (4), steering arms (4), **robotic ball device (4)**, motor axle (3), internal assembly (3), main casing (3), toy robot (3), internal magnetic element (3), feedback signals (3), **reactive actuation (3)**, more functions (3), forward motion (3), backward motion (3), power source (3), drive motors (3), top section (3), **remote controlled movable ball amusement device (3)**, drive unit (3), pivot axis (3), motor drive units (3), **transformable toy vehicle (3)**, oppositely disposed sides (3), central hub (3), circumferential surface portion (3), **animal likeness (3)**, driven wheel (3), steering motor (3), **ball surface (2)**, reference axis (2), **omniwheel assemblies (2)**, mobile robot (2), **spherical caps (2)**, control circuit (2), electric power source (2), external magnetic element (2), multimodal dynamic robotic systems robotic systems (2), **mechanical spinning robot toy (2), spherical body (2), solar cells (2)**, curved surface (2), its upright position (2), **sphere interior (2), hollow sphere (2)**, support surface (2), inner surface (2), **propulsion mechanism (2), spherical steering toy (2)**, mounting plate (2), inside wall (2), transmission gear train (2), **non-spherical ball (2), shell part (2)**, drive system (2), steering system (2), rotor portion (2), **cam surface (2)**, trailing end (2), upper housing cover (2), drive wheel assembly (2), individual vanes (2), vane axis (2), some embodiments (2), remote control (2), drive motor (2), first, second and third wheels (2), center point (2), forward linear direction (2), defined pathway (2), weighted component (2) |

Figure 4a: Comparing concepts extracted from initial and retrieved sets of patents.

In the initial set, many concepts (e.g., system controller) seem to indicate those that correspond to any robot. While the concepts as **highlighted**, e.g., self-propelling rolling toy, include original search keywords, in general, from the initial set, it seems hard to find those that offer inspiration from or learning of spherical rolling robots. On other hand, among the concepts in the retrieved set, many, as **highlighted**, e.g., spheroid shell, provide a surface-level understanding of spherical rolling robots. In addition, some useful inspiration (e.g., animal likeness) could also be found among these. It is also important to note that original search keywords – "rolling toy" or "rolling robot" are absent in the retrieved set.

### 4.2. Inventor's Portfolio (Design and Business)

We study the portfolio of an inventor named Pablo A. Valdivia Y. Alvarado who is a co-inventor of 35 patents. Using these patents as the initial set, we retrieve 36 patents. The purpose of this application is to examine whether the retrieved set of patents includes concepts that are similar to those in the portfolio, so as to identify if potential collaborators exist as inventors of the retrieved set of patents. In addition, retrieved concepts that lie in the domain of the portfolio, yet that offer inspiring ideas, could help identify new design and career opportunities. To compare the portfolio and the retrieved set of patents, we extract concepts as explained in the previous application and present these in Figure 4b.

| INVENTOR'S PORTFOLIO (N = 35) | RETRIEVED PATENTS (N = 36) |
|---|---|
| **screw member (32)**, **intervertebral cage (25)**, **mobile core (24)**, screw members (23), tapered end (20), threaded body (20), sliding box (20), **vertebral bodies (20)**, curved staple base (20), core rim (18), vertebral body (16), **internal angled screw guides (14)**, **internal screw guide (12)**, **vertebral endplates (12)**, concave trough (12), convex surfaces (12), bi-directional fixating transvertebral (11), **spinal fusion (10)**, **bi-directional fixating/locking transvertebral body (10)**, at least first and second sliding boxes (10), sliding boxes (10), self-drilling bone fusion screw apparatuses (10), cervical facet staple (10), at least two prongs (10), bottom surface (10), **bi-directional fixating transvertebral body screws (10)**, posterior cervical and lumbar interarticulating joint calibrated stapling devices (10), central screw locking lever (9), posterior lumbar intervertebral placement (8), anterior lumbar intervertebral placement (8), anterio-lateral thoracic intervertebral placement (8), anterior cervical intervertebral placement (8), spinal fusion a self-drilling bone fusion screw apparatus (8), (bdft) screw/cage apparatus (7), bdft apparatus (7), internal angled screw (7), posterior cervical and lumbar facet joint staple (6), artificial replacement disc (6), **substantially parallel plates (6)**, opposite sides (6), lateral movement (6), parallel plates (6), more insertion tools (6), replacement disc (6), bdft screws (5), **staple gun (5)**, cage stand (5), cage indentation (5), other screw (5), artificial cervical and lumbar discs (5), disc plate insertion gun (5), sequential single plate intervertebral implantation (5), symmetric bi-disc plate alignment (5), interplate mobile core placement (5), tool assembly (5), zero-profile expandable intervertebral spacer devices (4), universal tool (4), their placement (4), entire spine (4), anterior, anterolateral, lateral, far lateral or posterior surgical approaches (4) | screw members (31), lever arms (24), stapling apparatus (20), screw apparatus (20), spinal column (20), **vertebral bodies (20)**, opposite directions (16), **spinal fusion (15)**, **vertebral endplates (13)**, body portion (12), upper and lower disc plates (12), control device (12), distal end (12), self-drilling screw apparatus (10), building materials (10), **substantially parallel plates (10)**, **bi-directional fixating transvertebral body screws (9)**, zero-profile horizontal intervertebral miniplates (9), joint stapling device (9), **intervertebral cage (9)**, pinion shaft (9), upper and lower claws (8), total artificial expansile disc (7), drive mechanism (6), nail member (6), alignment slot (6), proximal end (6), artificial lumbar disc (6), **screw member (6)**, **internal angled screw guides (5)**, <u>expansile intervertebral body fusion devices (5)</u>, light emitting device (5), semi-circular rings (5), total intervertebral body fusion devices (4), worm drive screw (4), spur gear (4), superior and inferior screws (4), bi-directional manner (4), rotating mechanism (4), capping horizontal mini (4), grip handles (4), transmission linkages (4), drive rod (4), superior and inferior segments (4), <u>serrated interfaces (4), teethed unidirectional locking mechanism (4)</u>, four facet piercing elements (4), <u>staple/fuse adjacent spinous processes (4), thoracic/lumbar transverse process staples (4)</u>, staple/fuse adjacent transverse processes (4), <u>ratchet spring mechanism (4), bone fastener prongs (4)</u>, staple claw (4), adjacent spinous or transverse processes (4), adjacent processes (4), staple prongs (4), multiple perforations (4), bone fusion material (4), spinal elements (4), **bi-directional fixating/locking transvertebral body (4)**, expansion device (4), perimeter enclosure (4), **staple gun (4)**, puller tip (4), **mobile core (4)**, **internal screw guide (4)**, top claw (4), bottom claw (4) |

Figure 4b: Comparing concepts extracted from initial and retrieved sets of patents.

In Figure 4b, we **highlight** the concepts that are similar in both sets of patents. We also **highlight** the concepts that appear to be in-domain while also appear to offer design inspiration. Many **highlighted** concepts, e.g., "mobile core, bi-directional fixating/locking transvertebral body", are more frequent in the portfolio than in the retrieved set of patents, understandably because the retrieved set of patents is generalized within the domain. In addition, some concepts (e.g., spinal fusion) are more frequent in the retrieved set than in the portfolio, possibly representing the domain rather than the portfolio. This comparison allows us to infer the concepts that are domain-specific or portfolio-specific and better understand the portfolio of the inventor.

We also identify the concepts (e.g., teethed unidirectional locking mechanism) that appear to be within the domain but could offer inspiration for the inventor to develop new mechanisms or modify existing mechanisms. Such distant yet in-domain concepts are likely to be present in patents that do not share citation or inventor links with the initial set. The retrieval of such concepts is possible largely through patent embedding (as proposed in our method) than through query-based retrieval.

# Conclusions

The motive of our research is to enhance patent retrieval by devising an embedding approach for patent representation that captures multiple facets (e.g., text, inventor, citation) of knowledge that is communicated through patent document. We individually obtained the embeddings of these using Sentence-BERT (for text) and TransE (citation and inventor networks) and identified a suitable mode of combination through the performance in a classification task, which suggests that concatenation of text, citation, and inventor embeddings offer the preferred patent representation scheme. We also identified using a recall task that mean cosine similarity could be utilised for associating multiple initial patents and a target patent.

The application of the method with two sets of patents – product family and inventor's portfolio, suggests that the patents retrieved using our method include several concepts that have the potential to offer valuable design inspiration, instigate learning, indicate collaborations, and hint career opportunities. To realize the potential in terms of design and business applications, the method is preferably embodied in a support tool or integrated with existing patent search tools like InnoGPS[25] [17], [24].

---

[25] http://www.innogps.com/